\documentclass{osa-article}

\journal{oe}
\begin{document}

\title{Relative Intensity Noise in a Multi-Stokes Brillouin Laser}

\author{Ananthu Sebastian, Irina V. Balakireva, Schadrac Fresnel, St\'{e}phane Trebaol and Pascal Besnard \authormark{*}}

\address{\authormark{} Univ Rennes, CNRS, Institut FOTON - UMR 6082, F-22305 Lannion, France}

\email{\authormark{*}pascal.besnard@enssat.fr} %% email address is required



\begin{abstract}
We investigate the Relative Intensity Noise (RIN) properties of a multi-Stokes Brillouin fiber ring laser. We experimentally analyse intensity noise of each Stokes waves and study the noise dynamics of the cascaded Brillouin scattering process. We observe up to 20 dB/Hz intensity noise reduction compared to that of the RIN input pump laser. We examine the impact of the fiber ring quality factor on the laser RIN features such as amplitude reduction and relaxation frequency.  A numerical model based on a set of coupled-mode equations replicate the experimental observations; confirming the class B like behavior of a multi-Stokes Brillouin laser. Our study enables to determine the optimal parameter values to operate the multi-Stokes laser in the low noise regime.
\end{abstract}

%%%%%%%%%%%%%%%%%%%%%%%%%%  body  %%%%%%%%%%%%%%%%%%%%%%%%%%

\section*{\label{sec:level1}Introduction\protect\\}
Stimulated Brillouin Scattering (SBS) process \cite{Ippen1972} generate considerable interests in laser physics communities for its ability to induce low noise lasers \cite{Smith1991,Debut2000}.  Linewidth narrowing and intensity noise reduction have been demonstrated in various systems as fiber rings \cite{Geng2006}, whispering gallery mode microresonators \cite{Lin2014} or planar waveguide structures \cite{Buettner2014} to only name a few. Those systems are made of nonlinear materials as silica \cite{Molin2008}, chalcogenide \cite{Tow2012}, silicon nitride \cite{Gundavarapu2018}, crystal fluoride \cite{Grudinin2009} or silicon \cite{Shin2013}. SBS low noise lasers are attractive candidates for a large panel of applications ranging from coherent optical communications \cite{Choudhary2017}, RF signal generation\cite{Li2013} and processing \cite{Liu2018} to sensors \cite{Zarinetchi1991}. Laser performances have been widely studied and in particular intensity \cite{Stepien2002,Geng2006} and frequency noises \cite{Loh2016,Gundavarapu2018} reaching even metrological performances \cite{Loh2015, Suh2017}. Cascading the SBS process by using one Stokes order, as a pump, for the next red shifted Stokes wave brings the possibility to produce a comb source with equally spaced lines \cite{Lim1998}.\\
A remaining question concerns the potential of cascading the SBS process to improve the intensity and frequency noise reduction. Recently, a theoretical paper predicts that intensity and frequency noise reduction are not possible by cascading the SBS process \cite{Behunin2018}. It should be noted that noise reduction has been already observed in  Brillouin laser when second Stokes order starts to lase \cite{Tow2012, Dennis2010}. In this paper, we demonstrate that, under a specific operating regime, Stokes line intensity noise can be reduced up to 40 dB/Hz respect to the input pump RIN. Moreover, we report the complex RIN dynamics of a multi-Stokes Brillouin laser (MBL) in various configurations. In particular, we investigate the role of the ring cavity Q-factor on the relaxation of oscillation and on the RIN reduction of the MBL. Our numerical simulations support our experimental observations.\\
The paper is organized as the following: in Section \ref{sec:Experimental Details} we describe the experimental setup and the fiber ring resonator used for the present study. In Section \ref{sec:TheoreticalModel}, we introduce the MBL model based on coupled-mode equations. In Section \ref{sec:Cascading effect}, we derive the stationary response of MBL and recall its working principle. In Section \ref{sec:RIN}, we study the RIN behavior of Stokes lines in function of the input pump power and determine the conditions to obtain low-intensity noise laser emission. We finally discuss the role of cavity lifetimes on the RIN performance of the MBL.
\begin{figure}%[position]
\includegraphics[width=14cm]{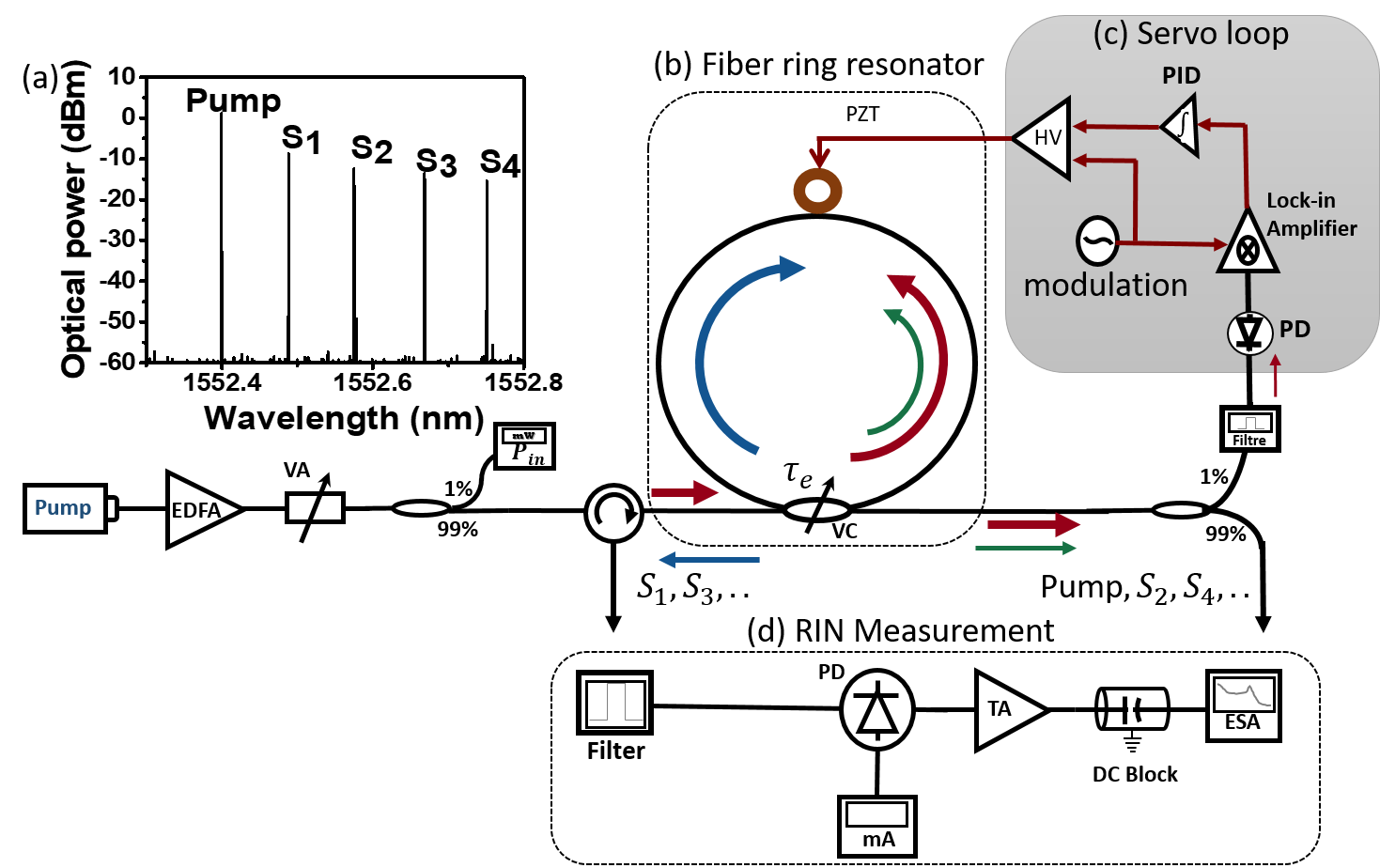}
\caption{Experimental setup to study the lasing properties of multi-Stokes Brillouin Laser. Pump: Koheras continous wave laser, EDFA: Erbium-doped fiber amplifer, VA: variable attenuator , VC: Variable coupler, PZT: Piezoelectric transducer, Filter: Yenista optical filter, PD: Photodiode,  PID: Proportional-integral differential amplifier, HV: High-voltage amplifer, TA: Transimpedance Amplifier, ESA: Electrical spetrum analyzer. The fiber ring cavity is composed of 20 m polarization maintaining fiber spooled around a PZT. Red line: Pump wave, Blue: Stokes 1 wave , Green: Stokes 2 wave.}\label{FigStaionnary}
\end{figure}

\section{\label{sec:Experimental Details}Experimental Details\protect\\}
\subsection{\label{FiberRingCavity}Fiber Ring Cavity}
The central element of the MBL is composed of a fiber ring resonator depicted in Fig. 1 (b). The gain or losses inside the resonator is quantified by the intrinsic cavity lifetime ($\tau_0/2$) and the coupling strength by the coupling lifetime ($\tau_e/2$). The total photon lifetime of the resonator is expressed as $1/\tau = 1/\tau_0 + 1/\tau_e$ bringing to the Q-factor equation $Q=\omega\tau/2$. The resonator is made of a 20 m polarization-maintaining fiber corresponding to a free-spectral range (FSR) of 10 MHz. The light is coupled in and out through a variable coupler allowing to vary the coupling coefficient and then the Q-factor of the resonator \cite{Dumeige2008}. In the present study, we consider two resonator configurations which we call: the high-Q resonator ($Q_\text{h}=6\times 10^8$) and the low-Q resonator ($Q_\text{l}=2\times 10^8$). Cavity parameters are extracted using the cavity ring down method \cite{Dumeige2008}. The low-Q factor and high-Q factor cavities are in the overcoupling ($\tau_0>\tau_\text{e}$) and the undercoupling ($\tau_0<\tau_\text{e}$) regimes respectively. In terms of Brillouin laser performances, the overcoupling regime implies a higher output power at the price of higher lasing threshold ($P_\text{th}= 26.5$ mW) with respect to the undercoupling operation ($P_\text{th}= 7.2$ mW). As it will be shown later, the Q-factor of the resonator impacts the RIN features of the MBL (Sec. \ref{sec:Impact of cavity parameters}).
\subsection{Experimental setup}
The experimental setup is illustrated in Fig. \ref{FigStaionnary}. The fiber ring cavity is optically pumped using a continuous wave (CW) 1 kHz line-width laser emitting at 1552.4 nm. The laser output power is amplified before the cavity injection. A motorized variable optical attenuator helps to control the input power in the MBL. 1\% of the input pump power is collected for power monitoring.\\
The laser pump signal is introduced in the cavity via a circulator and a coupler, and then propagates clockwise in the cavity (red arrow). The circulating pump wave initiates Brillouin amplification in Stokes 1 wave traveling in the opposite direction (blue arrow). For sufficiently high pump power ($P_\text{in}=P_\text{th}$), Brillouin lasing threshold is reached giving rise to efficient Stokes 1 emission outside the resonator. Stokes 1 wave can, in turn, play the role of pump laser for Stokes 2 wave (green arrow), the cascading process is then activated.
\subsubsection*{Cavity servoing}
The laser threshold for the various Stokes waves can be drastically reduced when the pump signal is resonantly coupled in the cavity. This configuration can be achieved by locking a cavity resonance to the CW pump laser. We implement a derivative spectroscopy stabilization method \cite{Weel2002}, its setup is presented in the gray box in Fig. \ref{FigStaionnary} (c). The actuator consists in a piezoelectric (PZT) ceramic cylinder wrapped by few meters of fiber. Varying the voltage on the piezoelectric implies a modification of the cavity length and then a control of the frequency position of the cavity resonances.\\
We apply a fast modulation frequency of 10 kHz and a slow modulation of 5 Hz on the PZT through a high-voltage amplifier. The slow modulation scans the cavity resonances. High-frequency dithering at 10 kHz creates derivative error signal from a lock-in amplifier. Actuator signal is produced by sending the error signal to a PID controller. Using this technique the cavity is maintained in resonance with the pump laser during MBL operation.

\subsubsection*{RIN measurement}

The usual measurement method of intensity noise characterization consists in the acquisition of the Power Spectral Density (PSD) of the intensity noise of a laser line \cite{Cox1998}. We will, therefore, use an Electrical Spectrum Analyzer (ESA) to obtain the PSD of the electrical signal at the detection level. The intensity noise measurement bench is schematically described in the Fig. \ref{FigStaionnary} (d). The detection system consists of a photodiode with a bandwidth from DC to 1 GHz, a transimpedance amplifier (TA) with a variable bandwidth depending on the gain, but not exceeding 200 MHz,  and a "DC-Block" module with very low cut-off frequency (1 Hz), to remove the DC component of the electrical signal in order to avoid damaging of the ESA. In Section \ref{sec:RIN}, we report analysis of the MBL RIN measurements for different Stokes lines.

\section{\label{sec:TheoreticalModel}Theoretical model\protect\\}
To study theoretically the RIN of generated Stokes orders in the cascaded SBS process, we extend the model described by W. Loh \cal{et al.} \cite{Loh2015a} in the framework of single Stokes generation, to the MBL. The model, based on the coupled modes formalism \cite{Haus1984}, depicts the temporal dynamics of $2N+1$ coupled-mode equations where $N$ is the maximum number of Stokes waves under consideration:
\begin{eqnarray}
\frac{\partial A_0}{\partial t}\! &=&\!- \frac{1}{ \tau} A_0\! -\! i q_0 \omega_0 A_1 \rho_1\! +\! \sqrt{\frac{2}{\tau_{e}}} S e^{i \sigma_0 t}, \label{ModelPump}\\
 \frac{\partial A_{\eta}}{\partial t}\! &=&\!- \frac{1}{ \tau_{\eta}} A_{\eta}\!-\!i q_{\eta} \omega_{\eta} \left[ A_{\eta-1} \rho^*_{\eta} + \delta_{\eta\neq N} A_{\eta+1} \rho_{\eta+1} \right], \label{ModelA}\\
\frac{\partial \rho_{\eta}}{\partial t}\! &=&\! i \frac{\Omega_b^2\! - \!\Omega^2_{\eta}}{2 \Omega_{\eta}} \rho_{\eta} \!-\! \frac{\Gamma_b}{2}\rho_{\eta} \!-\! i p_{\eta} A_{\eta-1} A^*_{\eta} .
\label{Modelrho}
\end{eqnarray}

Here $A_{\eta}, \omega_{\eta}$, and $\tau_{\eta}$ are the amplitude, frequency and lifetime of the generated Stokes wave label $\eta$ (with $\eta\in[1, N]$). Eq. (\ref{ModelPump}) describes the dynamics of the pumped mode $A_0$, its angular frequency is $\omega_0=1.22\!\times\! 10^{15}$ rad/s. $\tau_{e}$ is the pump-resonator coupling time constant. $\rho_{\eta}$ and $\Omega_{\eta}$ are the amplitude and frequency of the density wave label $\eta$. $\Gamma_b=2 \pi\! \times\! 30\! \times \!10^6$ rad/s is the loss rate of the density wave and $\Omega_b = 2\pi\! \times\! 11.55\! \times \!10^9$ rad/s is the acoustic frequency where the SBS gain is maximum. These two parameters are experimentally extracted. We consider a constant Brillouin Stokes shift over the cascading process, which means, $\Omega_b=\Omega_{\eta}$ for all $\eta$.  Coefficients $q_{\eta}$ and $p_{\eta}$ are:
\begin{eqnarray} 
q_{\eta}=\frac{\gamma_e }{4 n_0^2 \rho_0} \Lambda_{\eta}, \;\; \text{and} \;\; p_{\eta}= \frac{\epsilon_0 \gamma_e n_0^2 \Omega_{\eta}}{4 v^2} \Lambda_{\rho_{\eta}},
\end{eqnarray}
where $\gamma_e=1.5$ is the electrostrictive constant, $\epsilon_0$ is the vacuum permittivity, $n_0=1.44$ is the refractive index for silica, $\rho_0=2200$ kg/m$^3$ is the equilibrium density of the material, $v$ is the velocity of the acoustic wave. In the present work, the resonator is composed of fiber. We can reasonably consider that modes are similarly confined in the fiber and then fix the value of mode overlaps \cite{Loh2015} $\Lambda_{\eta}$ and $ \Lambda_{\rho_{\eta}}$ to $1$. The amplitude of the pump wave is $S=\sqrt{P/(V_{ph} \epsilon_0)}$, where $P$ is the power of the laser (in watts) and $V_{ph}=1.56\times 10^{-9}$ m$^3$ is the optical mode volume of the fiber ring cavity.\\
The total lifetime $\tau_{\eta}$ of each Stokes wave can be found by $1/\tau_{\eta}=1/\tau_{0\eta} +1/\tau_{e\eta}$ where $\tau_{0\eta}$ is the intrinsic lifetime (cavity losses) and $\tau_{e\eta}$ is the coupling lifetime. For the reason of simplification, we consider the lifetimes equal for all waves, and we write $\tau_{\eta} =\tau$, $\tau_{0\eta}=\tau_0$, $\tau_{e\eta}=\tau_e$, $q_{\eta}=q$, and $p_{\eta}=p$. In this article, we study the case of zero detuning between cavity resonance and the laser wavelength, which means $\sigma_0=\omega_L - \omega_0=0$. Experimentally, this configuration is maintained by the cavity servoing. We will see in the following that this resonant pump configuration is a key factor to get good matches with the experimental results.

\subsubsection*{\label{sec:RIN_Simulation} RIN Simulation}

To calculate RIN, we introduce noise in the pump term of Eq. (\ref{ModelPump}) as $S_0 = S + f_r$. Here $f_r$ represents the fluctuations of the pump amplitude approximated by a Langevin white Gaussian noise source with $\langle f_r(t)f^*_r(t') \rangle=C \delta(t-t')$, where $C$ is the auto-correlation strength of $f_r$. $C$ is used as a fitting parameter and it disappears after the normalization of Stokes RIN to the input pump RIN. Numerical simulations correspond to time streams of 7 ms with a time resolution of 3 ns. The corresponding quite large number of events justify the use of a Gaussian distribution, instead of a Poissonian one through the central limit theorem \cite{Petermann1988}. Then we take the amplitudes fluctuations $\delta  |A_{\eta}|=|A_{\eta}|-|A_{\eta}|_{\text{S}}$, and we find their spectral densities $S^p_{\delta |A_{\eta}|}$ by converting them in the frequency domain and multiplying by the complex conjugate. RIN is finally determined by \cite{Loh2015}:
\begin{equation}
\text{RIN}=\frac{8S^p_{\delta |A_{\eta}|}}{|A_{\eta}|_{\text{S}}^2}
\end{equation}
RIN curves represent the result from the averaging over 20 computations.

\section{\label{sec:Cascading effect} Cascade effect in Brillouin Fiber Laser\protect\\}
\begin{center}
\begin{figure}%[position]
 \includegraphics[width=12cm]{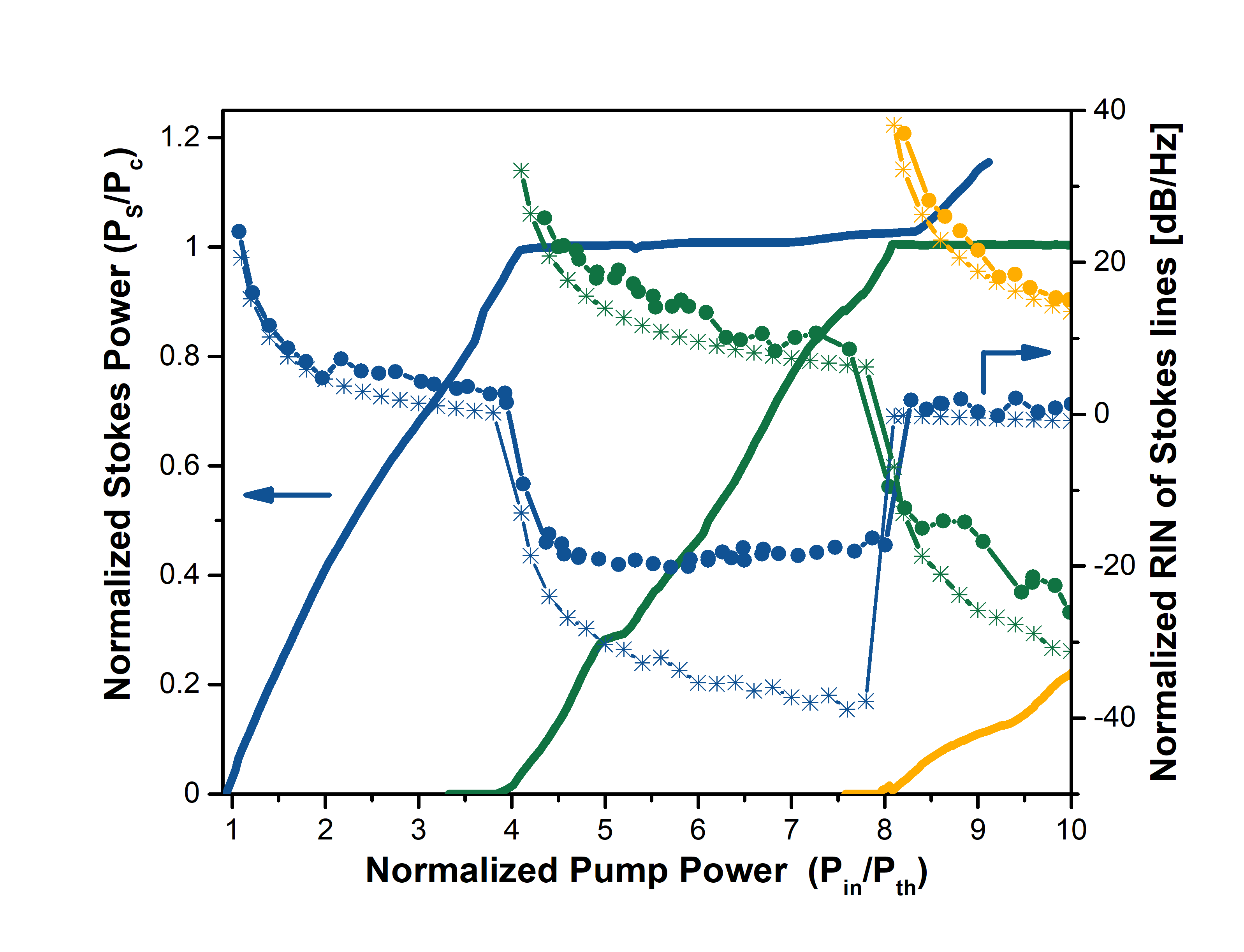}
 \caption{Left axis : Output Stokes power ($P_\text{s}$) normalized to the circulating power ($P_\text{c}$) versus input pump power ($P_\text{in}$) normalized to the Stokes 1 lasing threshold ($P_\text{th}$). Right axis: RIN of Stokes lines normalized to the input pump RIN. RIN amplitude is measured at 4 kHz from the carrier. Straight lines and full dots are experimental results. Stars are simulation results. Results are obtained with the "low-Q cavity" Brillouin laser, $P_\text{th}=26.5$ mW.}\label{fig:MultiStokes_RIN}
\end{figure}
\end{center}

We study the stationary response of a resonantly pump fiber cavity through the output power Stokes ($P_\text{s}$) as function of the input pump power. We recall the principle of the cascaded generation of Brillouin Stokes waves in such a resonator. We plot in Fig. \ref{fig:MultiStokes_RIN} (left axis), the evolution of output Stokes lines 1, 2, and 3 as function of the input pump power (straight lines) for the low-Q factor cavity. Independent Stokes lines output power characterization are performed by optical filtering.\\
Above the normalized pump power $P_\text{in}/P_\text{th}= 1$, the pump circulating power is clamped. This operating point corresponds to the onset of coherency in the first Stokes order. Indeed, the Brillouin gain, induced by the pump laser, balances the linear losses experience by Stokes 1 line: efficient stimulated Brillouin emission occurs (blue curve). When the power in Stokes 1 (S1) reaches the normalized output power needed to compensate S2 linear losses, coherent emission also takes place; the cascaded process is on. The cascaded generation of multiple Stokes waves has been treated analytically by Toyama et al. \cite{Toyama1993} considering resonant pumping of the cavity and constant Brillouin gain for the all Stokes orders. Analytical expressions for the Stokes power evolution versus input pump power are derived from Eqs. (\ref{ModelPump}, \ref{ModelA}, \ref{Modelrho}) (see Appendix). The model gives rise to similar results than previously reported models in the literature \cite{Toyama1993, Behunin2018}.\\
It is worth to mention that when the output power of a Stokes order ($\eta +1$) evolves monotonically with the pump power, the output power of the pumping Stokes order ($\eta$) is constant due to the clamping effect \cite{Ananthu2018}. Stokes lines output power alternate between monotonous evolution and constant power emission when the clamping effect occurs due to the onset of coherent emission on the $\eta +1 $ Stokes wave. This effect is determinant to explain the main contribution of the paper related to the intensity noise reduction of MBL discussed in the next Section.
\section{\label{sec:RIN} Relative intensity noise of multi-Stokes laser \protect\\}
This Section reports the main contribution of this paper regarding the intensity noise evolution of a multi-Stokes laser. We normalize the Stokes RIN level to the input pump RIN laser to determine the Brillouin laser RIN transfer function. In Section \ref{sec:RIN_VS_Cascade}, we scrutinize the overall evolution of the RIN for the Stokes waves $S1$, $S2$, and $S3$ as function of the input pump power. Then, in Section \ref{sec:RIN_VS_Stokesline}, we focus on the RIN behavior for individual Stokes lines $S1$ and $S2$, which are performed revealing the class B behavior of the multi-Stokes laser. In Section \ref{sec:Impact of cavity parameters}, we discuss the impact of the cavity Q-factor on the MBL RIN. All experimental RIN results are compared to numerical simulations performed with experimentally extracted parameters.
\subsection{\label{sec:RIN_VS_Cascade} RIN versus the cascade process}
As a first step in the study of RIN in MBL, we discuss the behavior of the RIN at 4 kHz from the carrier. We report in Fig. \ref{fig:MultiStokes_RIN} (right axis) the RIN behavior of the three first individual Stokes lines as function of the input pump power. Dotted lines correspond to experimental RIN measurements and starred lines to numerical simulations. We first focus on the $S1$ RIN. For a normalized input pump power of 1, corresponding to the $S1$ threshold, $S1$ RIN highlights 20 dB/Hz excess RIN with respect to that of normalized input pump. As mentionned before, the pump clamping between 1$\times P_\text{th}$ and 4$\times P_\text{th}$, implies a transfer of pump fluctuations towards $S1$ \cite{Loh2016}. Those fluctuations are amplified through the Brillouin gain explaining the excess noise. While going away from the threshold, the $S1$ output power increases implying a progressive decrease of the RIN towards the input pump RIN level. It is worth to notice that, in single mode Brillouin laser, the output Stokes RIN can even be lower than the input pump RIN. The noise reduction is strongly dependent of the cavity and Brillouin gain coefficients \cite{Stepien2002}, and the laser sensitivity to environmental perturbations as in a usual laser. In the present work, we take advantage of the SBS cascade effect to reduce Stokes noise level below the input pump RIN as described below.\\
Above $4\times P_\text{th}$, which corresponds to the $S2$ lasing threshold, the $S1$ experiences an abrupt RIN reduction up to $20$ dB compared to the input pump RIN level. This reduction originates from the clamping effect of $S1$ due to the appearance of the stimulated $S2$ emission. In that regime, the "gain=loss" condition acts as a driving force to maintain the $S1$ output power constant [See Eq. (\ref{StaticAN-1})]. Thereby, any intensity fluctuations of $S1$ are attenuated. At $8\times P_\text{th}$, coherent $S3$ emission arises. $S2$ becomes an efficient pump for $S3$, and similarly, intensity noise reduction occurs for $S2$ above the $S3$ lasing threshold power. Since $S2$ output power is clamped, $S1$ output power increases again proportionally to the input pump power. The consequences on the $S1$ RIN are instantaneous and manifest as an increase of noise level toward the input pump one. Concerning the $S3$ RIN, close to the threshold, its RIN reaches 36 dB/Hz in excess to the pump laser RIN and progressively shortens towards the input pump RIN level. This high RIN level can be attributed to the noise transfer through the cascade effect. Indeed, SBS process implies energy conservation between involved waves. Then, RIN of uphill waves are transfered, and even amplified, to downhill waves. Moreover, as mention by Behunin \cal{et al.} \cite{Behunin2018} spontaneous anti-Stokes emission can also degrade RIN of Stokes waves (not taken into account in our model).\\
Simulation procedure has been described in section \ref{sec:RIN_Simulation}. RIN behavior is very well reproduced by the numerical simulations (starred lines) in particular: (i) the monotonous reduction of RIN in free running regime and (ii) the sudden reduction in the clamping regime. A discrepancy in the RIN reduction amplitude between the model and the experiment exists in the clamping regime. The model expects up to -40 dB/Hz reduction when the experimental results tend to -20 dB/Hz. This difference is attributed to the limited sensitivity of our experimental bench.\\  
In this Subsection, we have shown the substantial impact of the clamping effect on the intensity noise properties of the laser. In other words, once the Stokes line labeled $\eta+1$ reaches the lasing threshold power, the "gain=loss" condition implies the clamping of the Stokes wave $\eta$ playing the role of the pump. In this regime, fluctuations of Stokes $\eta$ are sharply reduced, and up to 20 dB/Hz reduction compared to the RIN of the input pump laser is experimentally observed. Similar noise fluctuations reduction have been demonstrated in semiconductor optical amplifier (SOA) used in the saturation regime \cite{Sato2001,Danion2014}.\\
In the following Subsection, we describe in detail the RIN behavior of individual Stokes order for various operating points to show the rich dynamics of RIN when MBL emission takes place.
\begin{center}
\begin{figure}%[position]
\includegraphics[width=\textwidth]{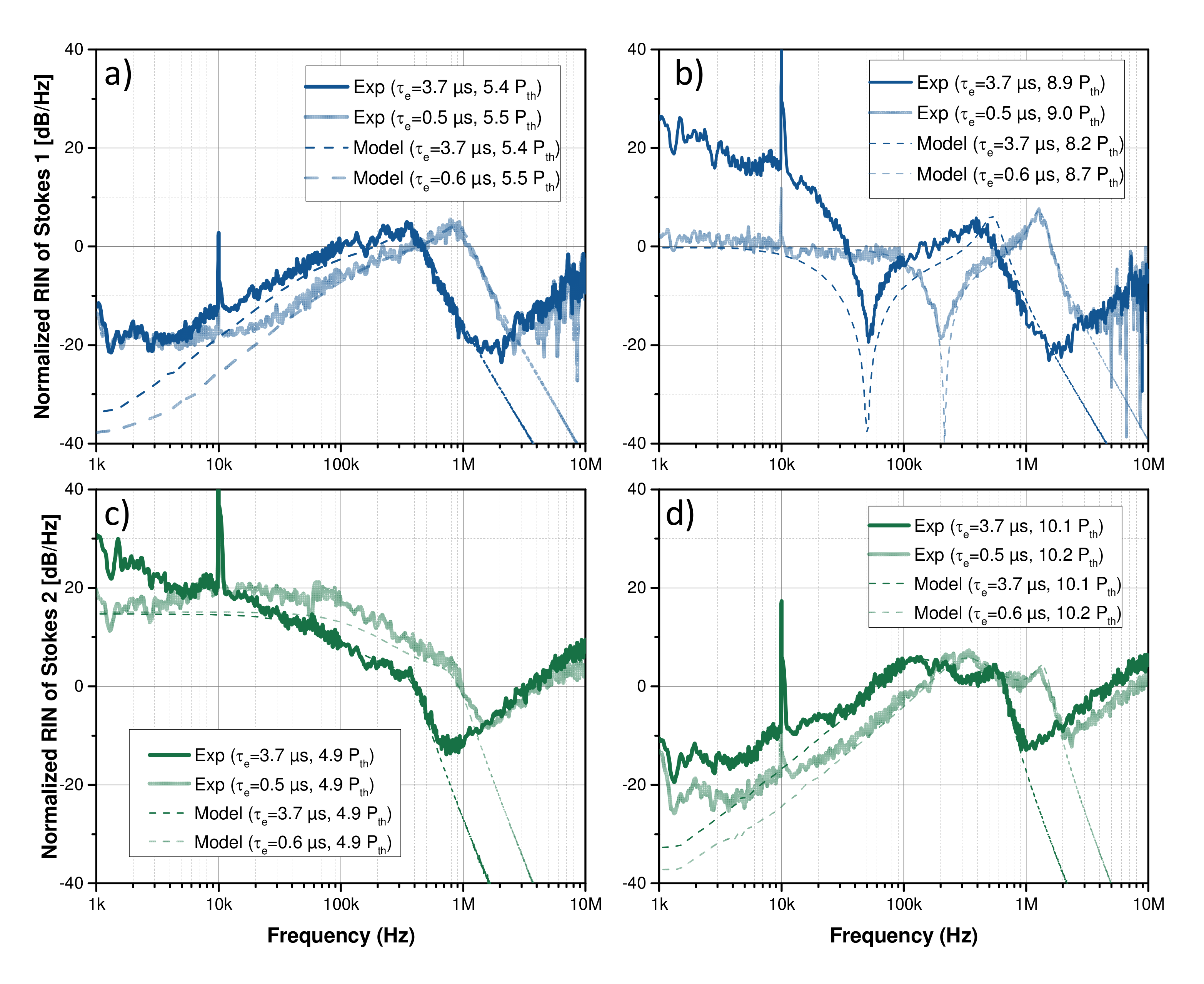}
 \caption{Normalized RIN of Stokes 1 (a, b) and Stokes 2 (c, d) lines for various input pump powers. Full and dashed lines hold for experimental and simulation results respectively. Deep and light colors refered to high ($\tau_0=1.4~\mu$s, $\tau_e = 3.7~\mu$s)  and low-Q factors ($\tau_0=1.4~\mu$s, $\tau_e = 0.5~\mu$s) respectively.}\label{fig:Stokes1_2_RIN}
\end{figure}
\end{center}
\subsection{\label{sec:RIN_VS_Stokesline} RIN of individual Stokes lines}
\begin{figure}%[position]
\includegraphics[width=11cm]{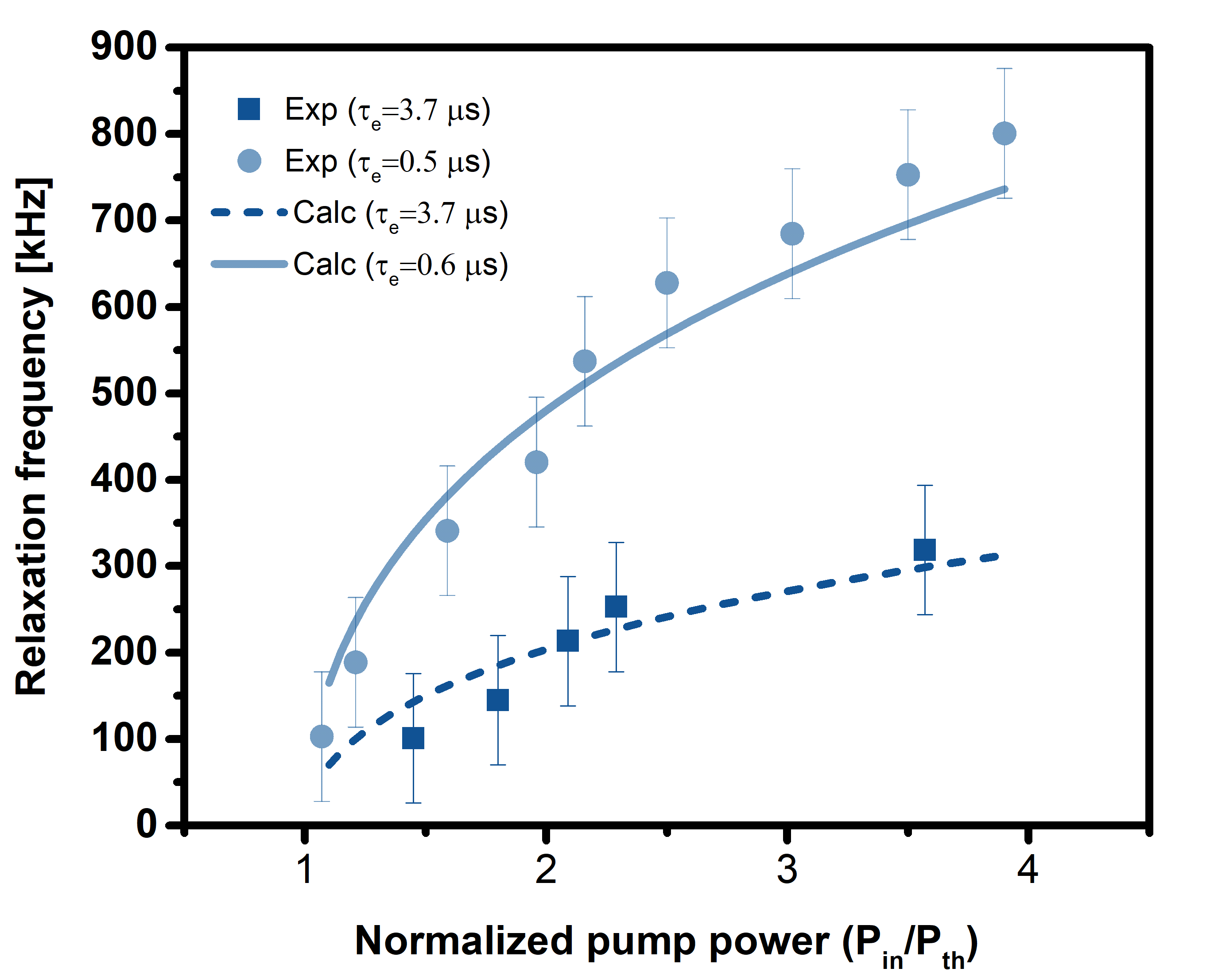}
 \caption{Relaxation frequency as function of the normalized pump power for low-Q (light blue) and high-Q (deep blue) cavities. Lines represent the results of calculation of Eq. (\ref{eq:relaxation_frequency}). Error bars ($\pm 75$ kHz) represent the average possible misreading of relaxation frequency on RIN curves.}\label{fig:relaxation_freq}
\end{figure}

To get more insight on the intensity noise of individual Stokes lines, we propose to study the complete RIN noise behavior for various input pump powers.\\
We plot in Fig. \ref{fig:Stokes1_2_RIN} the RIN of $S1$ (blue color) and $S2$ (green color) between 1 kHz and 10 MHz in two operating points: i) in clamping regime [Fig. \ref{fig:Stokes1_2_RIN} a) and d)]  ii) in monotonous regime [Fig. \ref{fig:Stokes1_2_RIN} b) and c)]. We will first focus on light blue curves corresponding to the low Q-factor cavity.\\
In the clamping regime [Fig. \ref{fig:Stokes1_2_RIN} a) and d)], a 20 dB reduction of the $S1$ and $S2$ RIN occurs at low offset frequency ($<100$ kHz). As explained previously, the onset of laser emission on $S_{\eta+1}$ implies the clamping of $S_\eta$ power and then the attenuation of its RIN amplitude. The RIN follows the usual noise distribution of a class B laser featuring a relaxation of oscillation. Its origin comes from the coupling between phonons and photons similarly to photons and electrons in a class B diode laser \cite{Loh2015a}. An analytical expression for the relaxation frequency in single Stokes regime ($P_\text{in}$ from 1 to 4 $\times P_\text{th}$) can be expressed as \cite{Loh2015a}
\begin{equation}
f_\text{R}=\frac{1}{2\pi}\sqrt{\frac{|A_1|^2}{|A_0|^2}*\frac{2 \Gamma_b}{\tau^2}\left(\frac{2}{\tau}+\frac{\Gamma_b}{2}\right)^{-1}}\label{eq:relaxation_frequency}
\end{equation}
Fig. \ref{fig:relaxation_freq} yields extracted frequency of relaxation from Stokes 1 in both cavity configurations (symbols). The results of calculations of Eq. (\ref{eq:relaxation_frequency}) are plotted in lines. The frequency of relaxation evolves proportionally to the square root of the pump term like a class B diode laser \cite{Agrawal2013}. Both cavities show an excellent superposition between the analytical expressions and experimental results.\\
%Implementing the parameter value for this cavity, we found a frequency of relaxation $f_\text{R} = xx Hz$ [Ref loh theo : might we fit the shift with Eq(28) ? $\rightarrow$ Ananthu].\\
%The relaxation of oscillation is still present as expected and show a frequency shift towards higher offset frequency proportionally to the $S1$ output power.
For $P_\text{in} > 4 \times P_\text{th}$, the laser shifts from single to multi-Stokes Brillouin lasing and then Eq. (\ref{eq:relaxation_frequency}) is not valid anymore. The onset of $S3$ line for $P_\text{in} > 8 \times P_\text{th}$ implies the appearance of multiple frequencies of relaxation originating from the complex Stokes waves coupling \cite{Behunin2018} as shown on the RIN of Stokes 2 [Fig. \ref{fig:Stokes1_2_RIN} d), inside the interval (100 kHz, 2 MHz)].\\
In the monotonous regime [Fig. \ref{fig:Stokes1_2_RIN} b) and c)], when the considered Stokes order has an output power that increases monotonically with the input pump power, RIN amplitudes at low offset frequency rise to at least the input pump RIN level. In this regime the noise transfer channel from "pump" to Stokes line is  open. It corresponds to noise transfer from input pump toward $S1$ in Fig. \ref{fig:Stokes1_2_RIN} b) or from $S1$ toward $S2$ in Fig. \ref{fig:Stokes1_2_RIN} c).\\
As shown in Fig. \ref{fig:Stokes1_2_RIN}, our experimental observations are well reproduced by numerical simulations based on the model exposed in Section \ref{sec:TheoreticalModel}. In particular noise reduction at low offset frequencies, complex relaxation frequency shape and position and noise amplitude are well reproduced. As expected, the model shows that the cavity lifetime and the pump power determine the frequency of relaxation resonances.\\
Similar RIN behavior is observed experimentally for incoming Stokes orders alternating sequences of low RIN level at low frequency offset when the Stokes line is clamped and high RIN level when the noise of previous Stokes order is efficiently coupled through SBS process.

%In the configuration iv), for an input pump power of 8.2 $P_\text{th}$ when $S3$ laser is switching on, resurgence of low frequency noise appears on $S1$ RIN with a low pass filter transfer function. Indeed, the presence of coherent $S3$ line implies the clamping of $S2$ line and then the recovery of a monotonous increase of the $S1$ output power. Here,  i). The $S1$ noise level is close to the one of the pump laser as just before the appearance of $S2$ lines (Blue curve on Fig \ref{fig:Stokes1_2_RIN} a)).\\
 
\subsection{\label{sec:Impact of cavity parameters} Impact of photon lifetimes \protect\\}

In this Section, we highlight the fundamental role of the cavity parameters on the RIN properties of the class B MBL. For this purpose, we vary the coupling coefficients to modify the coupling lifetime $\tau_e$ from $0.5~\mu$s  to $3.7~\mu$s. The intrinsic lifetime, $\tau_0$ of the resonator remains the same since no additional losses are introduced ($\tau_0=1.4~\mu$s). Details about the two configurations and their characterizations are given in Section \ref{FiberRingCavity}.\\
%\begin{figure}%[position]
% \includegraphics[width=9cm]{Pictures/Comp_Norm_S1_RIN_5_5pth.png}\\
% \includegraphics[width=9cm]{Pictures/Comp_Norm_S1_RIN_9pth.png}
% \caption{•}\label{fig:Comp_Norm_S1_RIN}
%\end{figure}
Fig. \ref{fig:Stokes1_2_RIN} reports the RIN behavior for two cavity configurations. The light and deep colored curves refer to the low-Q (shorter photon lifetime) and high-Q (longer photon lifetime) resonator configuration respectively. By comparing the RIN of both configurations, two main features can be associated with the total photon lifetime values.\\
First, the frequency of the relaxation resonance is directly determined by the phonon lifetime and the total photon lifetime of the cavity as shown by Eq. (\ref{eq:relaxation_frequency}). Indeed, the shorter lifetime cavity (low-Q) systematically highlights a larger relaxation frequencies for $S1$ and $S2$ lines. This behavior has been reported in Fig. \ref{fig:relaxation_freq}, showing the frequency redshift versus input pump power ($1$ and $4~\times P_\text{th}$) for both configurations.\\
Secondly, low offset frequency Stokes RIN levels are systematically higher for the high-Q resonator. As an example, the $S1$ RIN in the high-Q configuration (deep blue) presents an excess noise of 25 dB/Hz with respect to the low-Q configuration (light blue) [Fig. \ref{fig:Stokes1_2_RIN} b)]. A possible explanation can be found in the relative frequency fluctuations between the input pump laser and the cavity resonance. The steeper the cavity resonance is, the better the frequency fluctuations are transducted in intensity fluctuations. Then, the high-Q resonator RIN is more sensitive to thermal \cite{Loh2015} and mechanical noise \cite{Stepien2002} and favor the transfer of low frequency pump fluctuations.  
Those effects are not taken into account in the present model, which may explain the discrepancy in the RIN level observed at low offset frequency in Fig. \ref{fig:Stokes1_2_RIN}.
\section*{\label{sec:level1} Conclusion \protect\\}
In this paper, we demonstrate that low offset frequency RIN of considered Stokes lines can be strongly attenuated when operating in the clamping regime. This is particularly true for first Stokes orders where 20 dB/Hz RIN reduction is observed compared to the input pump RIN. Class B behavior of the MBL is revealed by our study of the impact of photon lifetime on the relaxation frequency. Experimental observations are well supported by numerical simulations that reproduced the main features of multi-Stokes RIN. Our study gives inputs to design optimized Brillouin fiber laser in term of RIN reduction. Our results suggest driving the Brillouin laser between 5 to 7 times the Stokes 1 threshold power. The fact that our results obtained in fiber ring resonators are normalized in terms of threshold power and input pump noise allows the transposition of our observations to other photonic structures as planar micro-cavity whatever the type of materials is used.
\section*{Funding}
The present work is supported under project FUI AAP20 SOLBO, with the help of BPI FRANCE and P\^{o}le Images $\&$ R\'{e}seaux.

\section*{Acknowledgments}
The authors would like to thank Sophie Larochelle,  Mohamed Omar Sahni and Rodolphe Collin for fruitfull discussions and Fr\'{e}d\'{e}ric Ginovart for his careful reading of the manuscript.

\appendix

\section*{\label{Stat_rep}Stationary response}
The stationary response can be found by setting all derivatives in Eqs.(\ref{ModelA}) \& (\ref{Modelrho}) to zero and by checking the joint solutions of the equations. The solutions can be found for any $N$ using three equations. Two of them do not depend on the parity of $N$:
\begin{eqnarray}
|A_{N-1}|^2&=&\frac{K_N}{\omega_{N}},\label{StaticAN-1}\\
|A_{\eta+2}|^2 &=&|A_{\eta}|^2 -\frac{K_{\eta+1}}{\omega_{\eta+1}}.
\label{StaticAA+2}
\end{eqnarray}
Another one is changing in dependence on the parity of maximum Stokes order $N$:
\begin{eqnarray}
\!\!\!\!|A_0|^2\!&=&\!\frac{2\tau^2}{\tau_{e}} |S|^2\! \times\! \left[ K^{-1} \omega_0 |A_1|^2 +1 \right]^{-2}\!, \; \text{if} \; N=2,4,6... \; \label{Even}\\
\!\!\!\!|A_1|^2\!&=&\!\left[ \frac{ \tau}{\sqrt{\tau_{e}/2}} \frac{|S|}{|A_0|} -1 \right] \frac{K}{\omega_0}, \qquad \text{if} \; N=1,3,5... \label{odd}
\end{eqnarray}
Here $K_{\eta}=\Gamma_b/(2 \tau p_{\eta} q)$. The technique of the calculations is the following. First, we calculate $|A_{N-1}|^2$ using Eq.~(\ref{StaticAN-1}), and after using Eq.~(\ref{StaticAA+2}) we find $|A_{N-3}|^2, |A_{N-5}|^2$ etc. until $A_0$ if $N$ is odd or $A_1$ if $N$ is even. After we use Eqs.~(\ref{Even}) or (\ref{odd}) and finish the missing lines by Eq.~(\ref{StaticAA+2}) until $|A_N|^2$.


%\bibliography{RIN_MS_V9}

\begin{thebibliography}{10}
\newcommand{\enquote}[1]{``#1''}

\bibitem{Ippen1972}
E.~Ippen and R.~Stolen, \enquote{{Stimulated Brillouin scattering in optical
  fibers},} {\protect\JournalTitle{Applied Physics Letters}} \textbf{21},
  539--541 (1972).

\bibitem{Smith1991}
S.~Smith, F.~Zarinetchi, and S.~Ezekiel, \enquote{{Narrow-linewidth stimulated
  Brillouin fiber laser and applications},} {\protect\JournalTitle{Optics
  letters}} \textbf{16}, 393--395 (1991).

\bibitem{Debut2000}
A.~Debut, S.~Randoux, and J.~Zemmouri, \enquote{{Linewidth narrowing in
  Brillouin lasers: Theoretical analysis},} {\protect\JournalTitle{Physical
  Review A}} \textbf{62}, 023803 (2000).

\bibitem{Geng2006}
J.~Geng, S.~Staines, Z.~Wang, J.~Zong, M.~Blake, and S.~Jiang, \enquote{{Highly
  stable low-noise Brillouin fiber laser with ultranarrow spectral linewidth},}
  {\protect\JournalTitle{IEEE Photonics Technology Letters}} \textbf{18},
  1813--1815 (2006).

\bibitem{Lin2014}
G.~Lin, S.~Diallo, K.~Saleh, R.~Martinenghi, J.-C. Beugnot, T.~Sylvestre, and
  Y.~K. Chembo, \enquote{{Cascaded Brillouin lasing in monolithic barium
  fluoride whispering gallery mode resonators},} {\protect\JournalTitle{Applied
  Physics Letters}} \textbf{105}, 231103 (2014).

\bibitem{Buettner2014}
T.~F. B{\"u}ttner, M.~Merklein, I.~V. Kabakova, D.~D. Hudson, D.-Y. Choi,
  B.~Luther-Davies, S.~J. Madden, and B.~J. Eggleton, \enquote{{Phase-locked,
  chip-based, cascaded stimulated Brillouin scattering},}
  {\protect\JournalTitle{Optica}} \textbf{1}, 311--314 (2014).

\bibitem{Molin2008}
S.~Molin, G.~Baili, M.~Alouini, D.~Dolfi, and J.-P. Huignard,
  \enquote{{Experimental investigation of relative intensity noise in Brillouin
  fiber ring lasers for microwave photonics applications},}
  {\protect\JournalTitle{Optics letters}} \textbf{33}, 1681--1683 (2008).

\bibitem{Tow2012}
K.~H. Tow, Y.~L\'{e}guillon, S.~Fresnel, P.~Besnard, L.~Brilland,
  D.~M\'{e}chin, D.~Tr\'{e}goat, J.~Troles, and P.~Toupin.,
  \enquote{{Linewidth-narrowing and intensity noise reduction of the
  2$^\text{nd}$ order Stokes component of a low threshold Brillouin laser made
  of Ge$_{10}$As$_{22}$Se$_{68}$ chalcogenide fiber},}
  {\protect\JournalTitle{Opt. Express}} \textbf{20}, B104--B109 (2012).

\bibitem{Gundavarapu2018}
S.~Gundavarapu, R.~Behunin, G.~M. Brodnik, D.~Bose, T.~Huffman, P.~T. Rakich,
  and D.~J. Blumenthal, \enquote{{Sub-Hz Linewidth Photonic-Integrated
  Brillouin Laser},} {\protect\JournalTitle{arXiv preprint arXiv:1802.10020}}
  (2018).

\bibitem{Grudinin2009}
I.~S. Grudinin, A.~B. Matsko, and L.~Maleki, \enquote{{Brillouin lasing with a
  CaF$_2$ whispering gallery mode resonator},} {\protect\JournalTitle{Physical
  review letters}} \textbf{102}, 043902 (2009).

\bibitem{Shin2013}
H.~Shin, W.~Qiu, R.~Jarecki, J.~A. Cox, R.~H. Olsson~III, A.~Starbuck, Z.~Wang,
  and P.~T. Rakich, \enquote{{Tailorable stimulated Brillouin scattering in
  nanoscale silicon waveguides},} {\protect\JournalTitle{Nature
  communications}} \textbf{4}, 1944 (2013).

\bibitem{Choudhary2017}
A.~Choudhary, M.~Pelusi, D.~Marpaung, T.~Inoue, K.~Vu, P.~Ma, D.-Y. Choi,
  S.~Madden, S.~Namiki, and B.~J. Eggleton, \enquote{{On-chip Brillouin
  purification for frequency comb-based coherent optical communications},}
  {\protect\JournalTitle{Opt. Lett.}} \textbf{42}, 5074--5077 (2017).

\bibitem{Li2013}
J.~Li, H.~Lee, and K.~J. Vahala, \enquote{{Microwave synthesizer using an
  on-chip Brillouin oscillator},} {\protect\JournalTitle{Nature
  communications}} \textbf{4}, 2097 (2013).

\bibitem{Liu2018}
Y.~Liu, A.~Choudhary, D.~Marpaung, and B.~J. Eggleton, \enquote{{Chip-Based
  Brillouin Processing for Phase Control of RF Signals},}
  {\protect\JournalTitle{IEEE Journal of Quantum Electronics}} \textbf{54},
  1--13 (2018).

\bibitem{Zarinetchi1991}
F.~Zarinetchi, S.~Smith, and S.~Ezekiel, \enquote{{Stimulated Brillouin
  fiber-optic laser gyroscope},} {\protect\JournalTitle{Optics letters}}
  \textbf{16}, 229--231 (1991).

\bibitem{Stepien2002}
L.~Stepien, S.~Randoux, and J.~Zemmouri, \enquote{{Intensity noise in Brillouin
  fiber ring lasers},} {\protect\JournalTitle{JOSA B}} \textbf{19}, 1055--1066
  (2002).

\bibitem{Loh2016}
W.~Loh, J.~Becker, D.~C. Cole, A.~Coillet, F.~N. Baynes, S.~B. Papp, and S.~A.
  Diddams, \enquote{{A microrod-resonator Brillouin laser with 240 Hz absolute
  linewidth},} {\protect\JournalTitle{New Journal of Physics}} \textbf{18},
  045001 (2016).

\bibitem{Loh2015}
W.~Loh, A.~A. Green, F.~N. Baynes, D.~C. Cole, F.~J. Quinlan, H.~Lee, K.~J.
  Vahala, S.~B. Papp, and S.~A. Diddams, \enquote{{Dual-microcavity
  narrow-linewidth Brillouin laser},} {\protect\JournalTitle{Optica}}
  \textbf{2}, 225--232 (2015).

\bibitem{Suh2017}
M.-G. Suh, Q.-F. Yang, and K.~J. Vahala, \enquote{{Phonon-Limited-Linewidth of
  Brillouin Lasers at Cryogenic Temperatures},} {\protect\JournalTitle{Physical
  review letters}} \textbf{119}, 143901 (2017).

\bibitem{Lim1998}
D.~Lim, H.~Lee, K.~Kim, S.~Kang, J.~Ahn, and M.-Y. Jeon, \enquote{{Generation
  of multiorder Stokes and anti-Stokes lines in a Brillouin erbium-fiber laser
  with a Sagnac loop mirror},} {\protect\JournalTitle{Optics letters}}
  \textbf{23}, 1671--1673 (1998).

\bibitem{Behunin2018}
R.~O. Behunin, N.~T. Otterstrom, P.~T. Rakich, S.~Gundavarapu, and D.~J.
  Blumenthal, \enquote{{Fundamental noise dynamics in cascaded-order Brillouin
  lasers},} {\protect\JournalTitle{Phys. Rev. A}} \textbf{98}, 023832 (2018).

\bibitem{Dennis2010}
M.~L. Dennis, P.~T. Callahan, and M.~C. Gross, \enquote{{Suppression of
  relative intensity noise in a Brillouin fiber laser by operation above
  second-order threshold},} in \emph{Photonics Society, 2010 23rd Annual
  Meeting of the IEEE,}  (IEEE, 2010), pp. 28--29.

\bibitem{Dumeige2008}
Y.~Dumeige, S.~Trebaol, L.~Ghi{\c{s}}a, T.~K.~N. Nguyen, H.~Tavernier, and
  P.~F{\'e}ron, \enquote{{Determination of coupling regime of high-Q resonators
  and optical gain of highly selective amplifiers},}
  {\protect\JournalTitle{JOSA B}} \textbf{25}, 2073--2080 (2008).

\bibitem{Weel2002}
M.~Weel and A.~Kumarakrishnan, \enquote{{Laser-frequency stabilization using a
  lock-in amplifier},} {\protect\JournalTitle{Canadian journal of physics}}
  \textbf{80}, 1449--1458 (2002).

\bibitem{Cox1998}
M.~Cox, N.~Copner, and B.~Williams, \enquote{{High sensitivity precision
  relative intensity noise calibration standard using low noise reference laser
  source},} {\protect\JournalTitle{IEE Proceedings-Science, Measurement and
  Technology}} \textbf{145}, 163--165 (1998).

\bibitem{Loh2015a}
W.~Loh, S.~B. Papp, and S.~A. Diddams, \enquote{{Noise and dynamics of
  stimulated-Brillouin-scattering microresonator lasers},}
  {\protect\JournalTitle{Physical Review A}} \textbf{91}, 053843 (2015).

\bibitem{Haus1984}
H.~Haus, \emph{{Waves and fields in optoelectronics}} (Prentice-Hall,, 1984).

\bibitem{Petermann1988}
K.~Petermann, \emph{{Laser {D}iode {M}odulation and {N}oise}} (Kluwer Academic
  Publishers, 1988).

\bibitem{Toyama1993}
K.~Toyama, S.~Huang, P.-A. Nicati, B.~Y. Kim, and H.~J. Shaw,
  \enquote{{Generation of multiple Stokes waves in a Brillouin fiber ring
  laser},} in \emph{Optical Fiber Sensors Conference,}  (1993).

\bibitem{Ananthu2018}
S.~Ananthu, S.~Fresnel, S.~Trebaol, F.~Ginovart, and P.~Besnard,
  \enquote{{Improvement of noise reduction in fiber Brillouin lasers due to
  multi-Stokes operation},} in \emph{Conference: Fiber Lasers and Glass
  Photonics: Materials through Applications,}  vol. 10683 Proc.SPIE, ed.
  (2018).

\bibitem{Sato2001}
K.~Sato and H.~Toba, \enquote{Reduction of mode partition noise by using
  semiconductor optical amplifiers,} {\protect\JournalTitle{IEEE Journal of
  Selected Topics in Quantum Electronics}} \textbf{7}, 328--333 (2001).

\bibitem{Danion2014}
G.~Danion, F.~Bondu, M.~Alouini \emph{et~al.}, \enquote{{GHz bandwidth noise
  eater hybrid optical amplifier: design guidelines},}
  {\protect\JournalTitle{Optics letters}} \textbf{39}, 4239--4242 (2014).

\bibitem{Agrawal2013}
G.~P. Agrawal and N.~K. Dutta, \emph{Semiconductor lasers} (Springer Science \&
  Business Media, 2013).

\end{thebibliography}

% Produces the bibliography via BibTeX.
\end{document}